# Investigating the Longitudinal Development of EAS with Ultra High Energies

**Abbas Rahi Raham, A. Al-Rubaiee\*, Majida H. Al-Kubaisy**

*Dept. of Physics, College of Science, Mustansiriyah University, Baghdad, Iraq*

## Abstract

The simulation of the extensive air showers was performed by investigating the longitudinal development parameters ($N$ and $X_{max}$) by using AIRES system version 19.04.0. The simulation was performed at the energy range ($10^{18}$-$10^{20}$ eV) for different primary particles (such as primary proton and iron nuclei) and different zenith angles. The longitudinal development curves of EAS are fitted using Gaussian function that gave a new parameters for different primary particles and different zenith angles at the energy range ($10^{18}$-$10^{20}$ eV).

## Introduction

Ultra-High Energy Cosmic Rays (UHECRs) are the most energies particles known in the Universe. The study of the cascades resulting from their interactions with atmospheric nuclei can provide an unique glimpse into hadronic interaction properties at center-of-mass energies more than one order of magnitude above those attained in human-made colliders [1]. The collision of a UHECR with an atmospheric nucleus initiates an extensive air shower of secondary particles developing in the traversed air mass, usually referred to a slant depth, $X$ [2].

Extensive air showers develop in a complex way as a combination of electromagnetic cascades and hadronic multi-particle production [3]. It is necessary to perform detailed numerical simulations of air showers to infer the properties of the primary cosmic rays that initiate them. But simulations are a challenge since the number of charged particles in a high energy shower can be enormous, perhaps exceeding $10^{10}$ eV [3]. The longitudinal development of EAS is depended on the energy and type of the incident primary particle [4]. One of its characteristics is the atmospheric depth of shower maximum ($X_{max}$), which is often used to reconstruct the elemental composition of primary cosmic rays [5, 6]. The number of charged particles (*N*) in EAS as a function of atmospheric depth is intimately related to the type and energy of the primary particle which can be simulated by AIRES code [7]. The simulation was performed for two primary particles (primary proton and iron nuclei) at the high energy range ($10^{18}$-$10^{20}$ eV) for different zenith angles (0°, 20° and 30°).

---

\*) For correspondence: Email: dr.rubaiee@uomustansiriyah.edu.iq

## The Longitudinal Profile

The shower longitudinal profile is a dependence of the shower particle number (*N*) on a given traversed atmospheric depth, *X*. The parameterization of shower profile commonly used in cosmic-ray experiments is Gaisser-Hillas formula [8]:

$$N(X) = N_{\max} \left(\frac{X}{X_{max}}\right)^{X_{max}/\lambda} \exp{(X_{max} - X)/\lambda}, \qquad (1)$$

where *X* is the atmospheric depth (in g / cm$^2$) ; $N_{max}$ is the number of charged particles; $X_{max}$ is the depth of shower maximum and $\lambda$ is a characteristic length parameter (a scale constant with a value 70 g /cm$^2$) [9]. The first interaction point is the location of initial collision of the cosmic ray particle with atmosphere . A forth parameter is often introduced in to Eq. (1), ostensibly to allow for a variable first interaction point [10]:

$$N(X) = N_{max} \left(\frac{X - X_o}{X_{max} - X_o}\right)^{(X_{max} - X_o)/\lambda} \exp{(X_{max} - X)/\lambda}, \qquad (2)$$

Where $X_0$ is the depth of the point of first interaction. The value of $X_0$ depends on the collision cross section and hence the energy and mass composition of the particle. The $X_{max}$ depends on the position of $X_0$, the shower energy and composition. But one can express the shower longitudinal development as function of the shower age *s* that defined as $s=3X/X+2X_{max}$ instead of depth *X*, by using *s*. Translating the depth $X_{max}$ into age *s* and using the normalized shower size $n=N/N_{max}$ then Eq. (2) becomes [11]:

$$n(s) = \left(1 - \frac{(1-s)3T_m}{(3-s)(T_m - T_o)}\right)^{T_m - T_o} e^{3T_m}(1 - s/3 - s) \qquad (3)$$

Where $T_m = X_{max}/\lambda$ and $T_0 = X_0/\lambda$.

## Results and Discussions

The Aires simulation code was applied to study generation primary particles (such as primary proton and iron nuclei) induced air showers at the primary energy range ($10^{18}$-$10^{20}$ eV) and explores the longitudinal development. Although the simulation of air showers at the lowest energy is not a very realistic application, therefore to investigate the longitudinal development by comparing the Aires simulation with experimental results by using Gaussian function that gave a new parameters for different energy ranges ($10^{18}$ – $10^{20}$ eV) and different zenith angles. All simulations are performed by the AIRES code which obtained by using the thinning factor $10^{-7}$ relative thinning case. The parameterization of longitudinal development of showers that initiated in EAS was performed using Gaussian function that gave a new four parameters for different primary particles. This function is given as:

$$N(X) = \eta + \frac{\alpha e^{\frac{-4\ln(2)(x-\zeta_c)^2}{\delta^2}}}{\delta\sqrt{\frac{\pi}{4\ln(2)}}} \qquad (4)$$

Where $N$ is the number of particles in the shower as a function of the shower depth; $\eta$, $\alpha$, $\zeta_c$ and $\delta$ are the obtained coefficients for the longitudinal development in EAS (see Table 1).

Table (1) the coefficients of the Gaussian function (Eq. 4) that used to parameterize the AIRES simulation for different primary particles, different energies and different zenith angles.

| Primary particle | Zenith angle | Energy (eV) | Coefficients | | | | $Ch^2$ | $R^2$ |
|---|---|---|---|---|---|---|---|---|
| | | | $\eta$ | $\zeta_c$ | $\delta$ | $\alpha$ | | |
| P | 0° | $10^{18}$ | -72933 | 927.592 | 421.514 | $4.19\times10^9$ | $5.52\times10^9$ | 0.9993 |
| | | $10^{19}$ | -7167433 | 829.2 | 647.316 | $6.09\times10^{10}$ | $6.97\times10^{10}$ | 0.998 |
| | | $10^{20}$ | $-3.2\times10^7$ | 874.277 | 575.727 | $4.75\times10^{11}$ | $4.53\times10^{13}$ | 0.999 |
| | 20° | $10^{18}$ | -675399 | 776.303 | 601.247 | $4.65\times10^9$ | $9.35\times10^9$ | 0.997 |
| | | $10^{19}$ | -9796814 | 797.139 | 637.833 | $6.53\times10^{10}$ | $1.53\times10^{12}$ | 0.998 |
| | | $10^{20}$ | $-5.8\times10^7$ | 822.561 | 634.056 | $4.86\times10^{11}$ | $4.45\times10^{13}$ | 0.999 |
| | 30° | $10^{18}$ | -1322534 | 782.077 | 617.335 | $9.06\times10^9$ | $2.7\times10^{10}$ | 0.998 |
| | | $10^{19}$ | -8283108 | 779.129 | 586.921 | $5.98\times10^{10}$ | $1.86\times10^{12}$ | 0.997 |
| | | $10^{20}$ | $-7.6\times10^7$ | 778.989 | 603.687 | $5.26\times10^{11}$ | $1.31\times10^{14}$ | 0.997 |
| Fe | 0° | $10^{18}$ | $-5.4\times10^7$ | 650.739 | 453.591 | $3.62\times10^{11}$ | $2.16\times10^{14}$ | 0.9953 |
| | | $10^{19}$ | $-2.3\times10^8$ | 721.357 | 420.679 | $3.32\times10^{12}$ | $1.08\times10^{16}$ | 0.9981 |
| | | $10^{20}$ | $-1.7\times10^8$ | 759.699 | 432.703 | $3.27\times10^{13}$ | $6.43\times10^{17}$ | 0.9988 |
| | 20° | $10^{18}$ | $-6\times10^7$ | 626.665 | 445.074 | $3.57\times10^{11}$ | $2.63\times10^{14}$ | 0.994 |
| | | $10^{19}$ | $-4.2\times10^8$ | 647.907 | 423.022 | $3.42\times10^{12}$ | $2.13\times10^{16}$ | 0.9957 |
| | | $10^{20}$ | $-1.8\times10^9$ | 735.811 | 416.83 | $3.13\times10^{13}$ | $7.29\times10^{17}$ | 0.9986 |
| | 30° | $10^{18}$ | $-5.6\times10^7$ | 959.99 | 404.781 | $3.35\times10^{11}$ | $4.78\times10^{14}$ | 0.9899 |
| | | $10^{19}$ | $-3.2\times10^8$ | 656.5 | 393.415 | $3.14\times10^{12}$ | $2.47\times10^{16}$ | 0.9953 |
| | | $10^{20}$ | $-2.5\times10^8$ | 692.305 | 404.334 | $3.04\times10^{13}$ | $1.35\times10^{18}$ | 0.9973 |

In Figure (1) it was shown the longitudinal development of primary proton that simulated using AIRES code for three high energies $10^{18}$, $10^{19}$ and $10^{20}$ eV for vertical and inclined showers.

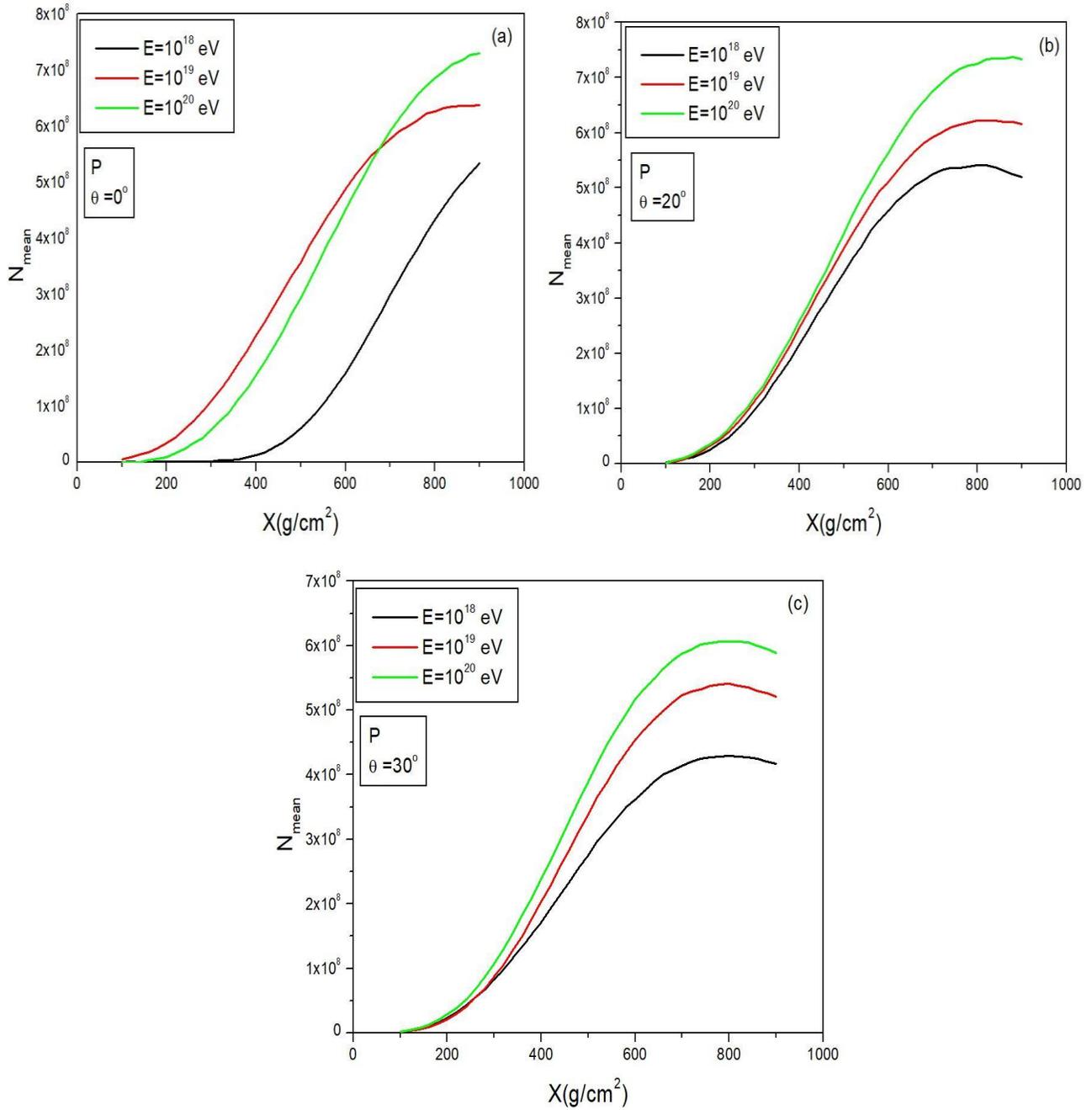

Figure (1): The longitudinal development simulated by AIRES code of primary proton for different zenith angles and different energies within the energy range ($10^{18}$-$10^{20}$ eV) for: (a) θ= 0˚; (b) θ= 20˚; (c) θ= 30˚.

Figure (2) reflects another importance or characteristic of the EAS development of the longitudinal development for iron nuclei that simulated by AIRES code for different zenith angles (0°, 20° and 30°) and different primary energies within the energy range ($10^{18}$-$10^{20}$ eV).

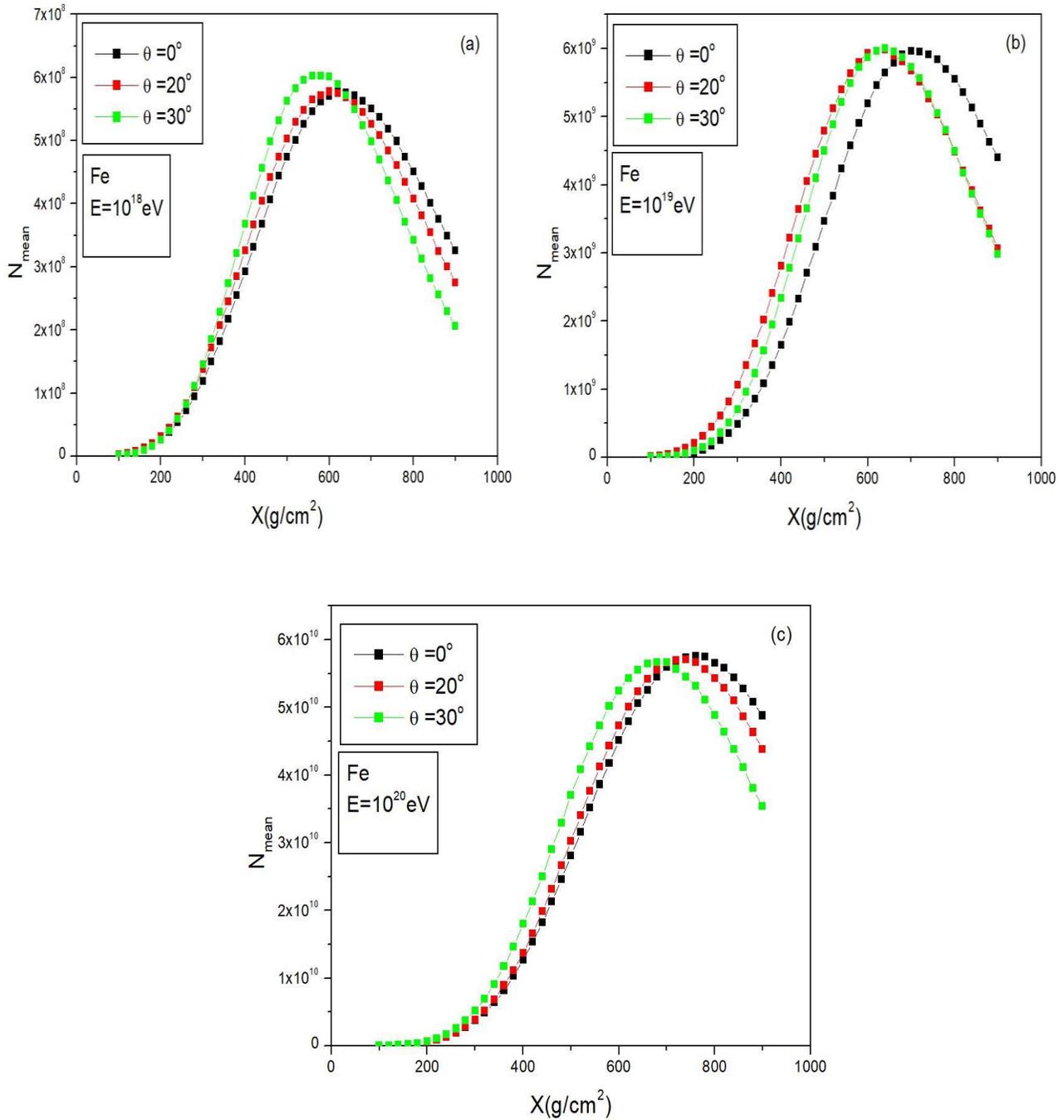

Figure (2): The longitudinal development that simulated by AIRES code for different zenith angles (0°, 20° and 30°) and different energies for iron nuclei at the primary energies: (a) $10^{18}$ eV; (b) $10^{19}$ eV; (c) $10^{20}$ eV.

Figure (3) displays the difference between the primary proton and iron nuclei of simulated longitudinal development for vertical showers. Through this figure one can see the effect of the energies on the fluctuations of the longitudinal development of primary particles for different primary energies.

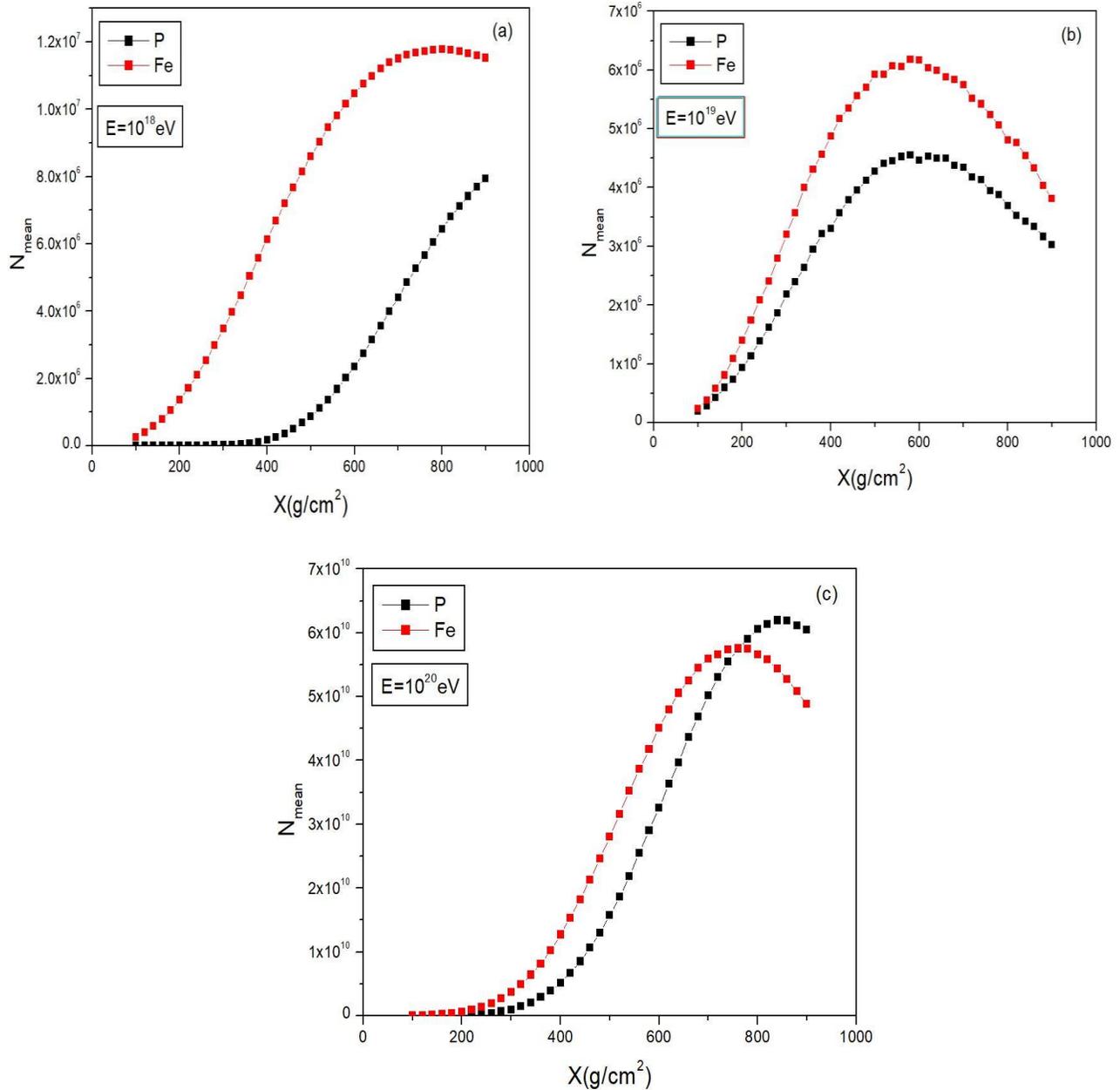

Figure (3): The longitudinal development that simulated by AIRES code for $\theta=0°$ and different energies for primary proton and iron nuclei at the primary energy: (a) $10^{18}$ eV; (b) $10^{19}$ eV ; (c) $10^{20}$ eV.

Through the figure (4) it was shown a good agreement in comparison between the parameterized longitudinal development using Gaussian function (Eq. 4) that gave a new parameters for different primary particles with the experimental measurements (AUGER experiment) [2, 12] for primary proton and iron nuclei at the energy $10^{19}$ eV for $\theta = 0°$.

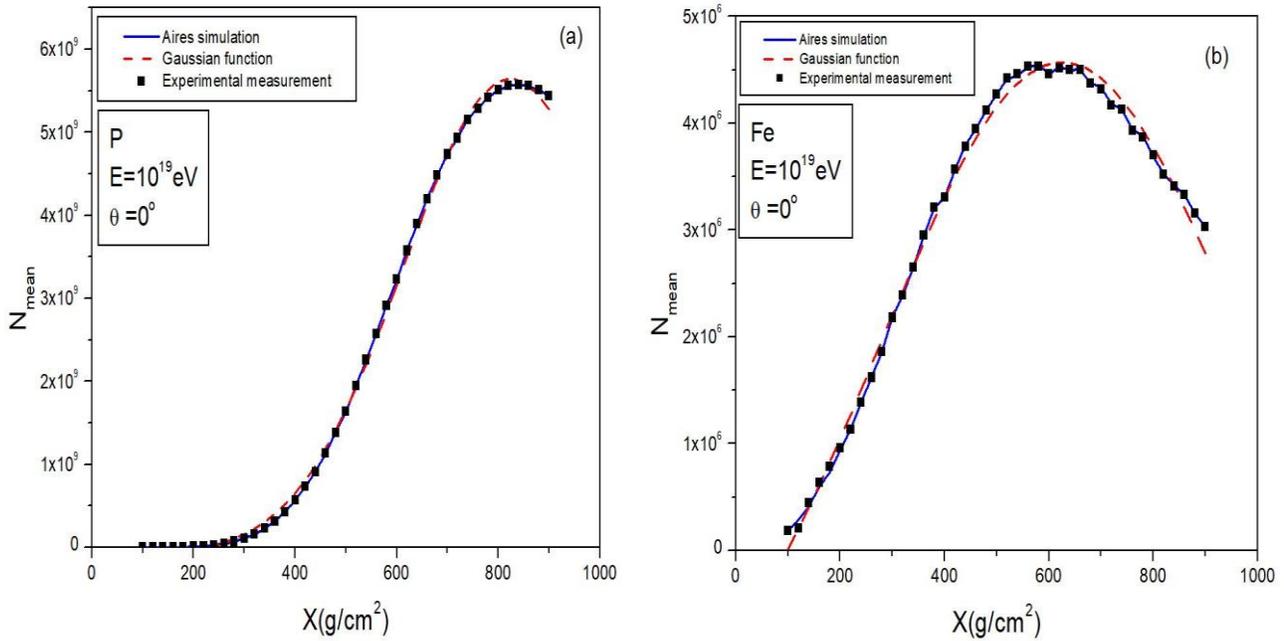

Figure (4) The comparison between the parameterization of longitudinal showers of AIRES simulation and the experimental measurements [2, 12] of vertical showers at fixed primary energy $10^{19}$ eV and different primary particles: (a) primary proton; (b) iron nuclei.

## Conclusions

The code AIRES was used for simulating the development of air showers in the atmosphere initiated by different primary particles (such as primary proton and iron nuclei) at the energy range ($10^{18}$-$10^{20}$eV) with different zenith angles. The simulation of the longitudinal development has shown an opportunity of primary particle identification and knowledge of the atmospheric depth for the cosmic ray spectrum. The dependence of the penetration depth at shower maximum, $X_{max}$, in extensive air showers as a function of the primary energy is considered within the high-energy range mentioned above. The parameters of the shower size (*N*) in EAS were obtained using the results of AIRES simulation as a function of the shower depth at maximum. The comparison of the parameterized longitudinal profile with that measured with the AUGER experiment demonstrates the ability for identifying the primary particles and to determine their properties at the very high energies around the ankle of cosmic ray spectrum. The main advantage of the given approach may give a possibility of longitudinal development analyzing for real events that detected in EAS arrays and for reconstructing the primary cosmic ray energy spectrum and mass composition.

**Financial Disclosure:** There is no financial disclosure.

**Conflict of Interest:** None to declare.

**Ethical Clearance:** All experimental protocols were approved under the Department of Physics/ College of Science/ Mustansiriyah University, Baghdad, Iraq and all experiments were carried out in accordance with approved guidelines.